\documentstyle[aps,twocolumn,graphicx]{revtex}

\begin{document}
\draft

\title{Magnetic storage device with improved temporal stability} 

\author{P.J. Jensen}

\address{Hahn-- Meitner-- Institut, Glienicker Str.100,
D -- 14 109 Berlin, Germany \\  e-mail: jensen@hmi.de}

\date{\today} 
\maketitle 

\begin{abstract}
The current efforts to fabricate non-volatile magnetic recording media
with a high  areal density is deteriorated by  the increasing temporal
instability  of the  stored  information.  If  the  stored energy  per
magnetic  particle  competes  with  the  thermal  energy,  spontaneous
magnetic  reversal processes  may  occur. Deposition  of the  magnetic
particles on  top of an antiferromagnetic substrate  will increase its
energy  barrier   due  to  the  exchange  coupling   between  the  two
subsystems. For  this the magnetic  moments of the  antiferromagnet in
the  vicinity of  the magnetic  particle  have to  deviate from  their
undisturbed  arrangement.   This disturbation  vanishes  within a  few
lattice  constants. In  the framework  of  a classical  spin model  we
calculate the  spin arrangements and the resulting energy barriers for
typical systems.  
\end{abstract} 
\vspace{0.5cm} 

A high areal density and  low noise of non-volatile magnetic recording
media is  achieved by use  of nanostructured thin films  consisting of
weakly  coupled  ferromagnetic   grains  deposited  on  a  nonmagnetic
substrate.   The  two  stable  magnetic states,  which  determine  the
information of a single bit,  are separated by an energy barrier which
is roughly proportional to the size of a single grain.  Currently much
effort  is being expended  on the  increase of  the areal  bit density
\cite{WeM99}. However, the decreasing bit size and thus the decreasing
energy  barrier   reduces  the   temporal  stability  of   the  stored
information.   Thermal  agitation   may  prompt  spontaneous  magnetic
reversal  processes  resulting  in  a  possible  loss  of  the  stored
information,   therefore  limiting   the  achievable   areal  density.
Technical requirements  demand a  lost of maximal  5 \% of  the stored
information  at  ambient  temperatures  over  10  years  \cite{CPY97}.
Considering  the known  magnetic  recording media  with high  magnetic
anisotropies and coercivities  such as Co/Pt multilayers \cite{Roo95},
and the current compound density  growth rate, this stability limit is
expected  to be reached  within a  few years.   To achieve  high areal
densities of  magnetic recording  media accompanied by  proper thermal
and temporal stabilities, we  propose that the required energy barrier
can be enhanced by depositing  the ferromagnetic grains or clusters on
top of an \em antiferromagnetic \em substrate.

The characteristic time to overcome the energy barrier is estimated in
the  framework  of the  Arrhenius-N\'eel  statistical switching  model
\cite{Nee49,Bro63}:
\begin{equation} 
\tau=\tau_0\;\exp(N\cdot\Delta E/k_BT)\;. \label{e1} 
\end{equation}
$N$ is the  number of atomic magnetic moments in  the cluster, $T$ the
absolute temperature,  and $k_B$  the Boltzmann constant.   The energy
barrier per cluster atom $\Delta  E$ results usually from the magnetic
lattice    anisotropy    and    the    magnetic    dipole    coupling.
$\tau_0^{-1}=10^9$ to $10^{12}$ sec$^{-1}$ is the attempt frequency to
overcome the  energy barrier.  The atomic  magnetic moments $\mu_{at}$
of a  cluster are assumed  to be ferromagnetically  ordered (collinear
magnetization), the cluster can thus be viewed to carry a single giant
magnetic   moment  $M=N\cdot  \mu_{at}$   (Stoner-Wohlfarth  particle)
\cite{StW48}.   The  above  mentioned temporal  stability  requirement
\cite{CPY97}  yields  the  ratio  $N\cdot\Delta E/k_BT\sim43$  at  the
least.

If   a  ferromagnetic   (FM)  cluster   is   placed  on   top  of   an
antiferromagnetic  (AFM)  substrate, the  FM  and  AFM subsystems  are
coupled by a (usually short range) interface exchange interaction.  If
due  to  this interface  coupling  the  magnetic  moments of  the  AFM
substrate close  to the FM cluster  are allowed to  deviate from their
equilibrium  (undisturbed) AFM  arrangement, the  total energy  of the
system  decreases  and  a   net  magnetic  coupling  between  the  two
subsystems emerges. Then the energy  barrier of the FM cluster between
its two stable states increases, and an enhanced temporal stability of
the  stored  information  is  obtained.   In  the  following  we  will
calculate the energy barriers for typical coupled FM -- AFM systems.

A  classical Heisenberg  Hamiltonian with  localized  magnetic moments
(spins)  {\boldmath$\mu$}$_i$ on  an fcc(001)  lattice  is considered.
The magnetic moments are  subject to the exchange interaction $J_{ij}$
between nearest  neighbors, the  lattice anisotropy $K_{i}$,  the long
range magnetic  dipole coupling, and the external  magnetic field {\bf
B} (Zeeman energy):
\begin{eqnarray}
E &=& -\frac{1}{2}\sum_{\langle i,j\rangle} \, J_{ij} \,
\mbox{\boldmath$\mu$}_i\,\mbox{\boldmath$\mu$}_j
-{\bf B}\,\sum_i\,\mbox{\boldmath$\mu$}_i-\sum_i\,K_i\,\cos^2\Phi_i 
\nonumber \\
&&+\frac{1}{2}\sum_{i,j \atop i\ne j}\frac{1}{r^5} \Big[
\mbox{\boldmath$\mu$}_i\,\mbox{\boldmath$\mu$}_j\,r^2 -3({\bf
r}\,\mbox{\boldmath$\mu$}_i ) ({\bf r}\,\mbox{\boldmath$\mu$}_j)
\Big]\,. \label{e2} \end{eqnarray}
The  FM and  the  AFM  subsystems are  characterized  by the  exchange
couplings  $J_{\rm FM}$  and $J_{\rm  AFM}$, the  anisotropies $K_{\rm
FM}$ and  $K_{\rm AFM}$, and  the magnetic moments $\mu_{\rm  FM}$ and
$\mu_{\rm AFM}$.   For these quantities typical values  are taken into
account. In addition the FM and  the AFM subsystems are coupled by the
interface  exchange interaction $J_{\rm  FM-AFM}$. This  coupling also
causes  the  unidirectional anisotropy  (exchange  bias) observed  for
extended FM-AFM  interfaces \cite{NoS99}. The distance  between a spin
pair is given by  $r=|{\bf r}|=|{\bf r}_j-{\bf r}_i|$.  For simplicity
the spins  are allowed to  rotate only in  the plane of  the substrate
face, with $\Phi_i$ the in-plane angle of the $i$-th spin.

The FM cluster with a  finite vertical and lateral extension is placed
on  top of an  AFM substrate.   Due to  the magnetic  interactions the
spins of  both the FM cluster  and the AFM substrate  may deviate from
their undisturbed directions (easy axes).  Only in the vicinity of the
FM cluster the AFM spins  will deviate markedly from their equilibrium
directions.  Thus, for  our calculations  a finite  region of  the AFM
substrate close to the FM cluster will be considered, in which the AFM
spins are  allowed to move. Its extension  has to be chosen  in such a
way that the  calculated energy barrier per spin  and other quantities
do not change if its range  is enlarged.  This disturbed AFM region is
embedded in an undisturbed extended AFM substrate

The energy barrier $\Delta E$ is calculated from the magnetic reversal
of the FM cluster. An  external magnetic field {\bf B} with sufficient
strength is  applied, forcing the  FM cluster magnetization  to rotate
 from one to its other easy direction.  For  given coupling parameters
and magnetic field angle $\Phi_B$ the directions $\Phi_i$ of each spin
of the  FM cluster and the  disturbed AFM region are  varied until the
total energy  $E(\Phi_B)$, equation (\ref{e2}), of  the coupled system
is minimal.  This  energy minimum needs not necessarily  be the global
minimum.   The Zeeman energy  has to  be subtracted  from $E(\Phi_B)$.
The  corresponding energy barrier  $\Delta E$  is determined  from the
difference between the minimum and the maximum of $E(\Phi_B)$.

In  Fig.1  a typical  behavior  of  $E(\Phi_B)$  in units  of  K/spin,
diminished  by  the Zeeman  energy,  is shown  as  a  function of  the
magnetic  field angle  $\Phi_B$.  A  square FM  cluster with  a single
atomic layer and  a lateral extension of about  five lattice constants
is assumed.  The range of the disturbed AFM region is choosen to be 12
lattice  constants in  lateral and  10 lattice  constants  in vertical
direction.  For the coupling constants  and magnetic moments of the FM
cluster  we use  $J_{\rm FM}=160$  K, $K_{\rm  FM}=-0.5$  K, $\mu_{\rm
FM}=0.6\;\mu_B$, and for the  AFM system $J_{\rm AFM}=-130$ K, $K_{\rm
AFM}=0.5$ K, $\mu_{\rm AFM}=2.5\;\mu_B$.  These values are typical for
a  bulk Ni ferromagnet  and a  NiO antiferromagnet  \cite{LaBoe}.  The
interface  coupling is  chosen  to  be $J_{\rm  FM-AFM}=7$  K in  this
example.  As  can be seen from the  full line in Fig.1,  this value of
$J_{\rm FM-AFM}$ refers to an energy barrier $\Delta E$ which is twice
the  energy  barrier  $\Delta  E_0$  of  the  decoupled  case  $J_{\rm
FM-AFM}=0$ (dotted  line).  $\Delta E_0$  is determined mainly  by the
lattice  anisotropy $K_{\rm  FM}$ of  the FM  cluster.  A  doubling of
$\Delta E$ increases the  characteristic time of the magnetic reversal
and thus the temporal stability by a factor e$^2\sim8$.

In Fig.2 we show the resulting energy barrier $\Delta E$ as a function
of the interface coupling  $J_{\rm FM-AFM}$. Four different values for
the  exchange  interaction $J_{\rm  AFM}$  of  the  AFM substrate  are
assumed,  refering  to  different  substrate  materials.   As  can  be
estimated  by  a  simple  calculation,  $\Delta  E$  increases  almost
quadratically with  $J_{\rm FM-AFM}$. Furthermore, for  the same value
of $J_{\rm  FM-AFM}$ the  energy barrier is  the smaller  the stronger
$J_{\rm  AFM}$. A  stronger  deviation  of the  AFM  spins from  their
equilibrium arrangement  increases the magnetic binding  energy to the
FM cluster.

\begin{figure}[t]  \label{Fig1} \hspace*{-0.5cm} 
\includegraphics[width=9cm,height=6cm,bb=30 440 520 775,clip]
{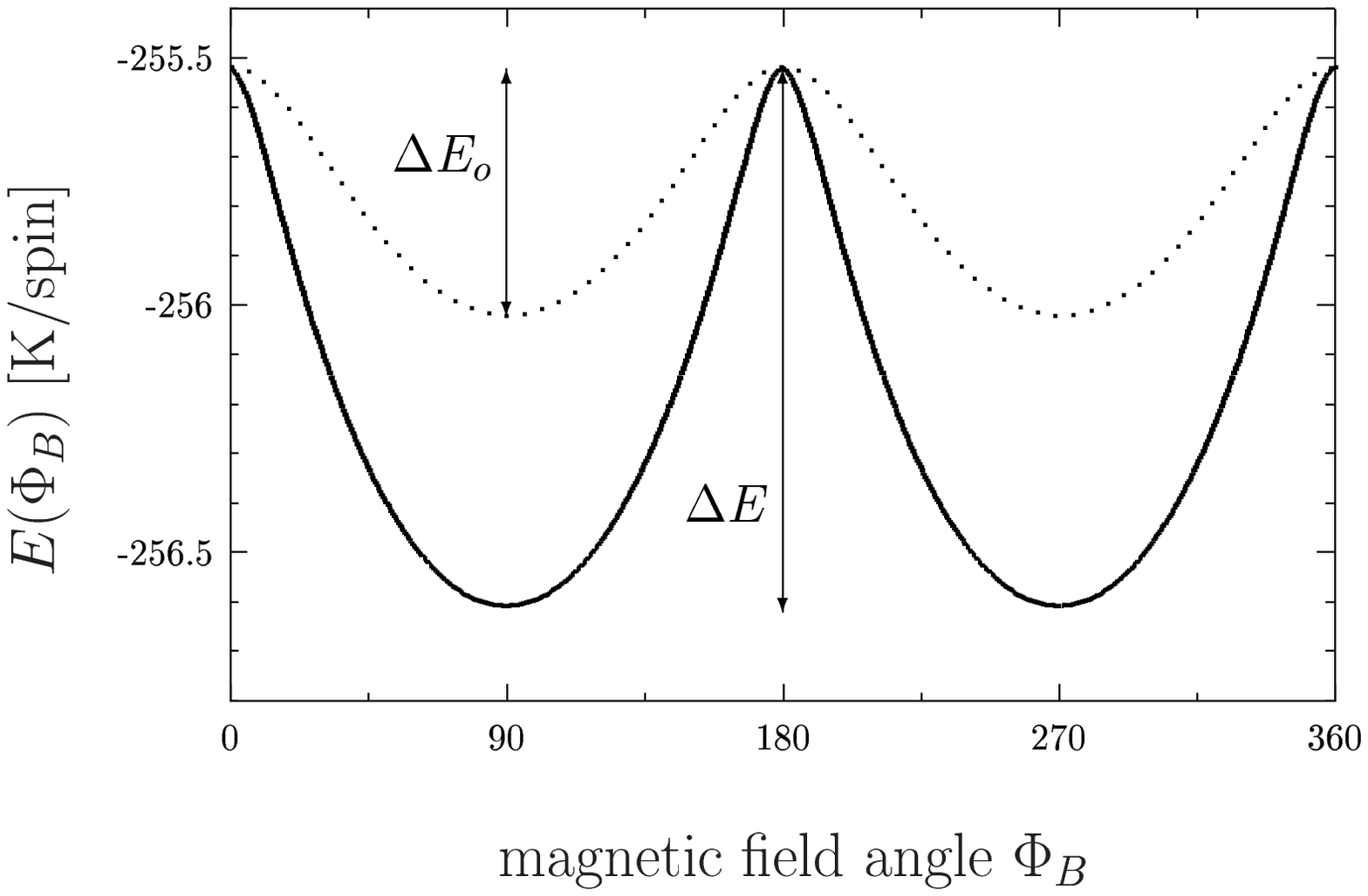}

\caption{Magnetic energy  $E(\Phi_B)$  in units  of
K/spin  as  a function  of  the direction  $\Phi_B$  (in  deg) of  the
external magnetic  field.  The Zeeman energy has  been subtracted from
$E(\Phi_B)$. The full line refers  to an interface coupling of $J_{\rm
FM-AFM}=7$ K, and  the dotted line to a  vanishing interface coupling.
$\Delta  E$ and $\Delta  E_o$ are  the corresponding  energy barriers.
For the values of the other coupling constants we refer to the text.}
\end{figure}

\begin{figure}[t]  \label{Fig2} \hspace*{-0.5cm}
\includegraphics[width=9cm,height=6cm,bb=30 460 500 780,clip]
{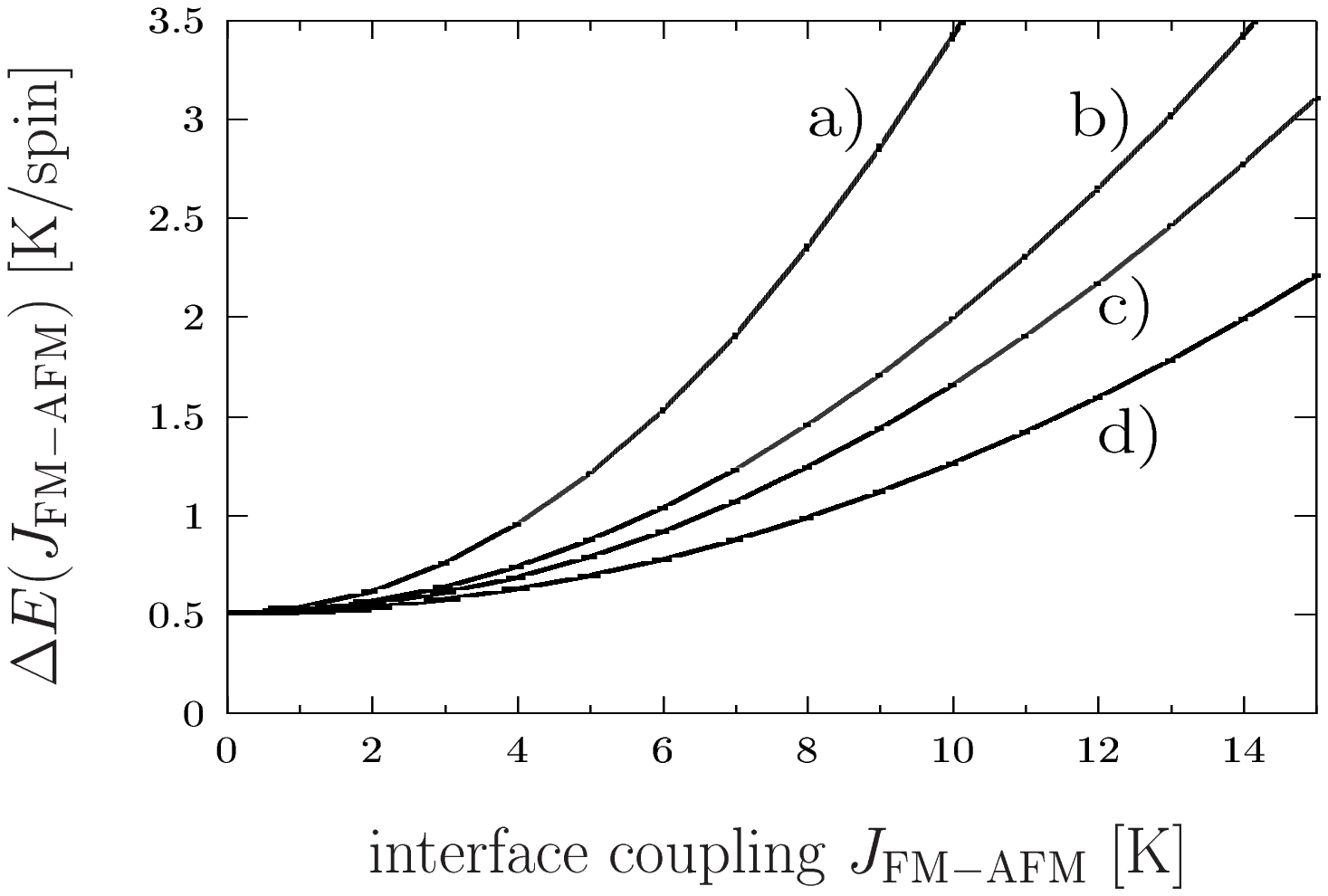}

\caption{Energy barrier $\Delta E(J_{\rm FM-AFM})$ in
units  of K/spin  as  a  function of  the  interface coupling  $J_{\rm
FM-AFM}$.  Four  different values of the  exchange interaction $J_{\rm
AFM}$ of the  AFM substrate are assumed: (a)  $J_{\rm AFM}=-50$ K, (b)
$J_{\rm  AFM}=-100$  K, (c)  $J_{\rm  AFM}=-130$  K,  and (d)  $J_{\rm
AFM}=-200$ K.  For the values of the other coupling constants we refer
to the text.} 
\end{figure}

We  have used  the interface  coupling  $J_{\rm FM-AFM}$  purely as  a
parameter.  With respect to  technical requirements a particular value
for  this  quantity  can  be  fixed by  choosing  a  certain  material
combination  of the coupled  FM --  AFM system.  For example  this may
enclose the  $3d$-transition- and the  $4f$-rare-earth-metals, as well
as  their alloys  and  compounds with  other  elements.  In  addition,
$J_{\rm FM-AFM}$ will depend on the atomic morphology of the interface
between  the FM  cluster and  the AFM  substrate.  We  note  that this
interface needs  not necessarily be  very flat, an  interface coupling
will  be  present also  for  rough  interfaces.   Also, the  interface
coupling  can   be  controlled  by  adding   magnetic  or  nonmagnetic
impuritites  or a spacer  layer between  the FM  clusters and  the AFM
substrat.   An oscillating  interlayer coupling  between ferromagnetic
layers as a  function of the spacer layer  thickness has been observed
for a number of multilayer systems \cite{HeB94}.

The enhanced  energy barrier due  to $J_{\rm FM-AFM}$ is  an interface
effect, since only the FM cluster  spins close to the interface to the
AFM substrate experience the interlayer coupling. The FM cluster spins
not located near the FM/AFM  interface are subject only to the lattice
anisotropy and the magnetic dipole  coupling, as for a cluster located
on a nonmagnetic substrate. Thus, assuming the same number of spins in
the  FM cluster,  the contribution  of the  interface coupling  to the
energy barrier is smaller for a  compact than for a flat island, since
the resulting interface  area is smaller for the  compact island. Thus
the  relative contribution of  the interlayer  coupling to  the energy
barrier   \em  per   spin   \em  will   decrease   for  a   decreasing
interface-to-volume ratio  of the cluster. By controlling  the size as
well as the shape of the FM clusters the energy barrier $\Delta E$ can
be varied according to technical requirements.

The reading  and in  particular the writing  process of  the recording
media must be performed independently bit by bit. This infers that the
FM  grains which carry  different informations  must be  placed within
such a  distance that  the writing  process of a  single bit  does not
interfere  with   the  state  of  its  neighboring   bits.   From  our
calculations we find that the disturbation of the AFM spin arrangement
decays  rapidly within  a few  lattice constants.  Both the  degree of
disturbance and its  range become larger if the  interface coupling to
the FM cluster increases.

The energy  barrier $\Delta E$  per spin of  the FM cluster  should be
large  enough to guarantee  the required  temporal stability.   On the
other hand, $\Delta E$  must not be too large in order  to allow for a
controlled  magnetic reversal  by an  external magnetic  writing field
with  appropriate strength and  within a  short switching  time.  This
implies  an upper  limit for  $\Delta E$  and thus  for  the interface
coupling.

The  above calculations have  been performed  at zero  temperature.  A
finite temperature $T$ can be  considered, if a free energy expression
with temperature dependent coupling coefficients is applied, resulting
in  a  temperature  dependent  energy barrier  $\Delta  E(T)$.   These
coefficients  can be  calculated  e.g.\ by  a  molecular field  theory
\cite{JeB98}.  Usually  a finite temperature  facilitates the reversal
of  the  cluster magnetization.   For  an  increasing temperature  the
characteristic time  to overcome  the energy barrier  (switching time)
becomes smaller, cf.\ equation (\ref{e1}).  Note that for temperatures
above its magnetic ordering temperature  a single FM cluster cannot be
viewed as a Stoner-Wohlfarth particle,  resulting in an upper limit of
the operation  temperature of such granular systems.   The coupling to
the extended AFM  system will improve the magnetic  ordering of the FM
cluster at finite temperatures, in particular for small clusters.

Acknowledgement: Numerous discussions with H.-D.Hoffmann are gratefully
acknowledged.

\end{document}